\documentclass[preprints,accept,moreauthors,pdftex,10pt,a4paper]{mdpi}
\firstpage{1} 
\pubvolume{xx}
\issuenum{1}
\articlenumber{1}
\pubyear{2018}
\copyrightyear{2018}
\history{Received: date; Accepted: date; Published: date}

\Title{Dynamical Casimir effect and state transfer in the ultrastrong coupling regime}

\Author{Giuliano Benenti $^{1,2,3}$*\orcidA{}, Michele Stramacchia $^{1,2}$ and Giuliano Strini $^{4}$}

\AuthorNames{Giuliano Benenti, Michele Stramacchia and Giuliano Strini}

\address{%
$^{1}$ \quad Center for Nonlinear and Complex Systems, Dipartimento di Scienza e Alta Tecnologia, Universit\`a degli Studi dell``Insubria, via Valleggio 11, 22100 Como, Italy\\
$^{2}$ \quad Istituto Nazionale di Fisica Nucleare, Sezione di Milano, via Celoria 16, 20133 Milano, Italy\\
$^{3}$ \quad NEST, Istituto Nanoscienze-CNR, I-56126 Pisa, Italy\\
$^{4}$ \quad Department of Physics, University of Milan, via Celoria 16, 20133 Milano, Italy}

\corres{Corresponding author: giuliano.benenti@uninsubria.it}

\abstract{The dynamical Casimir effect (DCE) manifests itself in the ultrastrong matter-field coupling (USC) regime, 
as a consequence of the nonadiabatic change of some parameters of a system. We show that the DCE is a 
fundamental limitation for standard quantum protocols based on quantum Rabi oscillations, implying that new schemes
are required to implement high-fidelity ultrafast quantum gates. Our results are illustrated by means of a paradigmatic 
quantum communication protocol, i.e., quantum state transfer.}


\begin{document}


\section{Introduction}

High-speed manipulation of quantum systems is vital for the prospects of quantum 
information processing. In quantum computation, quantum gates should operate on a time scale 
much smaller than the decoherence time to allow fault-tolerant architectures. 
In quantum communication, achieving high transmission rate is fundamental to 
boost applications of quantum cryptography. 
Circuit quantum electrodynamics (cQED) might play a prominent 
role to speed up quantum protocols, 
since it allows one to experimentally address the USC regime 
of light-matter interaction, where the coupling strength $g$ becomes comparable to, 
or even exceeds the resonator frequency 
$\omega$ \cite{lupascu,semba}. 

On the other hand, the performance of quantum protocols operating in the USC regime
might be significantly deteriorated, even in the absence of decoherence mechanisms, 
by the DCE \cite{casimirqip}. DCE is the generation of photons from the vacuum
due to time-dependent boundary conditions or, more generally, to the 
nonadiabatic variation of some 
parameters of a system 
(for reviews on the DCE see \cite{dodonov,noriRMP}, while 
a list of recent papers investigating various aspect of this phenomenon includes
\cite{ciuti,jaskula,exotic,koghee,solano2014,frigo,johansson2015,adesso2015,savasta2015,DCEtomo,DCEoptimal,macri18,DCEpert}). 
This latter case is expected to be relevant in the 
quest for ultrafast quantum protocols, requiring ultrafast driving of a quantum system.   

Here, we demonstrate the strong impact of photon emission
by the DCE on a paradigmatic quantum protocol, that is, state transfer
from qubit 1 ($\textsf{Q}_1$)
to qubit 2 ($\textsf{Q}_2$), mediated by a single mode of the quantized
electromagnetic field (cavity mode $\textsf{C}$).
The cavity acts as a quantum-bus,
which allows to reliably move quantum information and share entanglement
between different units of a quantum computing 
architecture \cite{blais,wallraff,sillanpaa,falci}.
 
\section{Model}

The qubits-cavity dynamics is described by the
quantum Rabi Hamiltonian~\cite{micromaser,qcbook,QRM}, with switchable couplings:
\begin{equation}
  \begin{array}{c}
{\displaystyle
H(t)=H_0+H_I(t),
}
\\
{\displaystyle
H_0=-\frac{1}{2}\,\sum_{k=1}^2\omega_k \sigma_z^{(k)} +
\omega\left(a^\dagger a +\frac{1}{2}\right),
}
\\
{\displaystyle
H_I(t)=\sum_{k=1}^2f_k(t)\,[\,g_k \,\sigma_+^{(k)}\,(a^\dagger+a)
+g_k^\star \sigma_-^{(k)}\,(a^\dagger+a)\,],
}
\end{array}
\label{eq:noREWAquantum}
\end{equation}
where we set $\hbar=1$,
$\sigma_i^{(k)}$ ($i=x,y,z$) are the Pauli matrices for qubit $\textsf{Q}_k$
($k=1,2$),
$\sigma_\pm^{(k)} = \frac{1}{2}\,(\sigma_x^{(k)}\mp i \sigma_y^{(k)})$
are the rising and lowering operators for the two-level system:
$\sigma_+^{(k)} |g\rangle_k = |e\rangle_k$,
$\sigma_+^{(k)} |e\rangle_k = 0$,
$\sigma_-^{(k)} |g\rangle_k = 0$,
$\sigma_-^{(k)} |e\rangle_k = |g\rangle_k$;
the operators $a^\dagger$ and $a$ create
and annihilate a photon:
$a^\dagger |n\rangle=\sqrt{n+1}|n+1\rangle$,
$a |n\rangle=\sqrt{n}|n-1\rangle$,
$|n\rangle$ being the Fock state with $n$ photons.
The switching on/off of the couplings is governed by the functions
$f_k(t)$, in the manner detailed below.
For simplicity's sake, we
consider the resonant case ($\omega_1=\omega_2\equiv\omega$) and
the coupling strengths $g_1=g_2\equiv g\in\mathbb{R}$.
The rotating-wave approximation (RWA) is obtained when we neglect the terms
$\sigma_+^{(k)} a^\dagger$, which simultaneously
excites $\textsf{Q}_k$ and creates a photon,
and $\sigma_-^{(k)} a$, which de-excites $\textsf{Q}_k$ and
annihilates a photon. In this limit, Hamiltonian
(\ref{eq:noREWAquantum}) reduces to the Jaynes-Cummings
Hamiltonian \cite{micromaser,qcbook}, with a switchable coupling.
We set $\omega=1$, so that in the RWA the swapping time needed to transfer
an excitation from one qubit to the field or vice versa
($|e\rangle_k |0\rangle\leftrightarrow |g\rangle_k |1\rangle$)
is $\tau=\pi/2g$
and the vacuum Rabi frequency $\Omega=g$.
The RWA approximation is a good approximation when $g/\omega\ll 1$ but 
fails in the USC regime.
We work in the interaction picture, where
the effective Hamiltonian at resonance is given by
$\tilde{{H}}(t)=
e^{i{H}_0t} H_I(t) e^{-i{H}_0 t}$
(we will
omit the tilde from now on).

\section{Rabi-type state-transfer protocol}

In order to transfer a generic pure state $|\psi\rangle=\alpha|g\rangle+\beta|e\rangle$
from qubit $\textsf{Q}_1$ to qubit $\textsf{Q}_2$, we consider the following quantum
protocol, based on quantum Rabi oscillations. We first discuss the 
protocol within RWA, where the state transfer is exact. 
Initially, $\textsf{Q}_1$ is prepared in the state $|\psi\rangle$,
while $\textsf{Q}_2$ and the cavity mode $\textsf{C}$ are in their ground state.
Then $\textsf{Q}_1$ interacts with $\textsf{C}$, for
a time $\tau$, so that the cavity is at the end in the state 
$|\tilde{\psi}\rangle=\alpha|0\rangle-i\beta|1\rangle$
and $\textsf{Q}_1$ in $|g\rangle$. 
The coupling of $\textsf{Q}_1$ with $\textsf{C}$ is then switched off and
$\textsf{Q}_2$ interacts with $\textsf{C}$,
for a time $\tau$. As a result, the state 
of $\textsf{Q}_2$ is driven to 
$|\hat{\psi}\rangle=\alpha|g\rangle-\beta|e\rangle$,
while $\textsf{C}$ is left in $|0\rangle$.
The transfer of state $|\psi\rangle$ to qubit $\textsf{Q}_2$ is recovered
after a rotation through an angle $\pi$ about the $z$ axis of the 
Bloch sphere for that qubit.  
When the terms beyond the RWA are taken into account,
state transfer is no longer perfect and 
the final state of $\textsf{Q}_2$ is given by
\begin{equation}
\rho'={\rm Tr}_{\textsf{Q}_1\textsf{C}}
[U (|\psi\rangle_1{}_1\langle \psi|\otimes |0\rangle\langle 0|\otimes|g\rangle_2{}_2\langle g|)U^\dagger],
\end{equation}
with $U$ unitary time evolution operator
for $\textsf{Q}_1 \textsf{C} \textsf{Q}_2$,
determined by the above described
quantum protocol.
We consider sudden switch on/off of the couplings,
i.e. $f_1(t)=1$ for $0\le t\le  \tau$, $f_1(t)=0$ otherwise;
$f_2(t)=1$ for $\tau\le t \le 2\tau$, $f_2(t)=0$ otherwise.
The quality of state transfer is measured by the fidelity
\begin{equation}
F=\langle\psi|\rho'|\psi\rangle.
\end{equation}
In the Jaynes-Cummings limit, the state transfer is perfect and the fidelity $F=1$.

\section{Results}



\begin{figure}[H]
\centering
\includegraphics[width=7.5 cm]{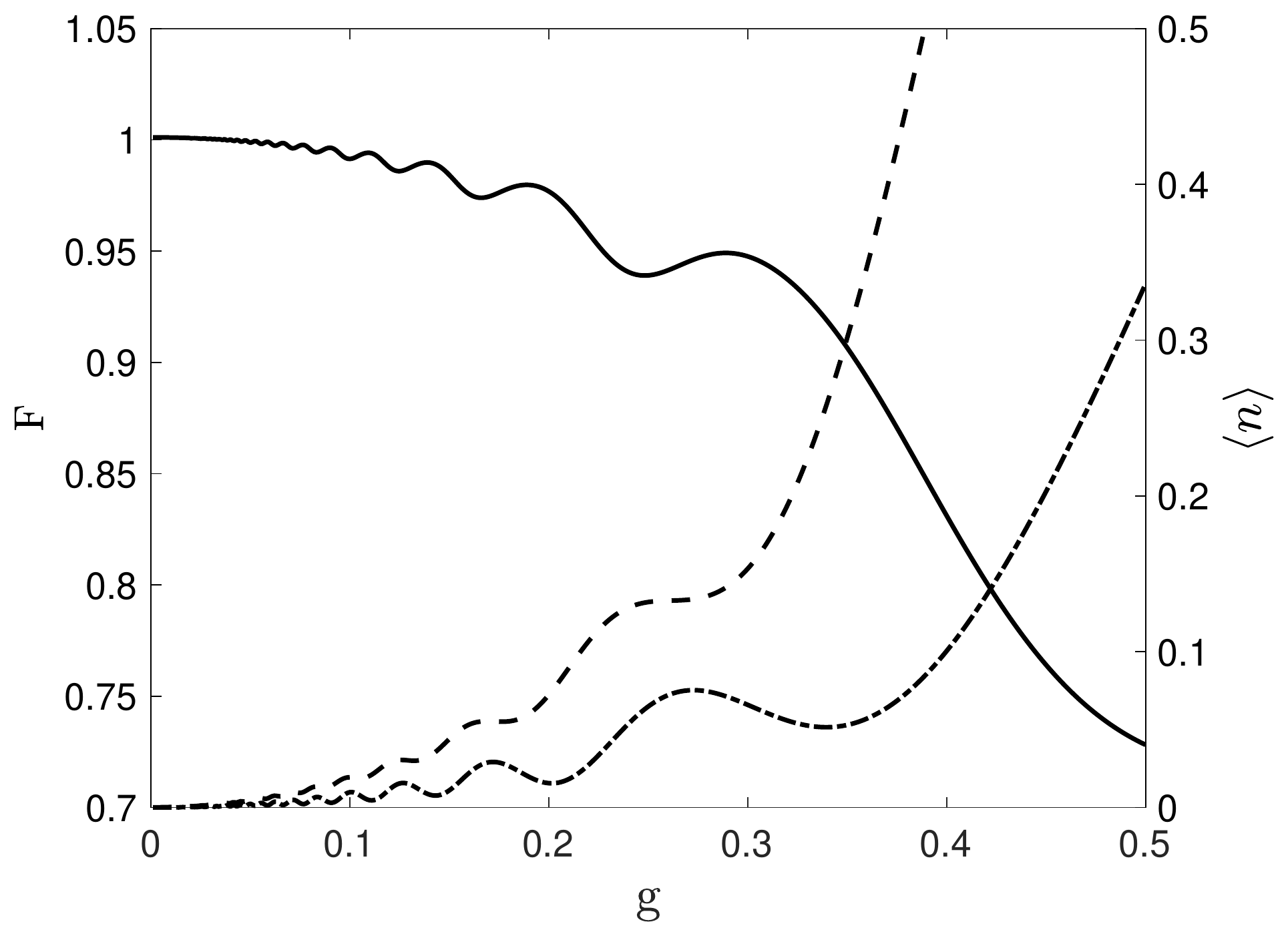}
\includegraphics[width=7.5 cm]{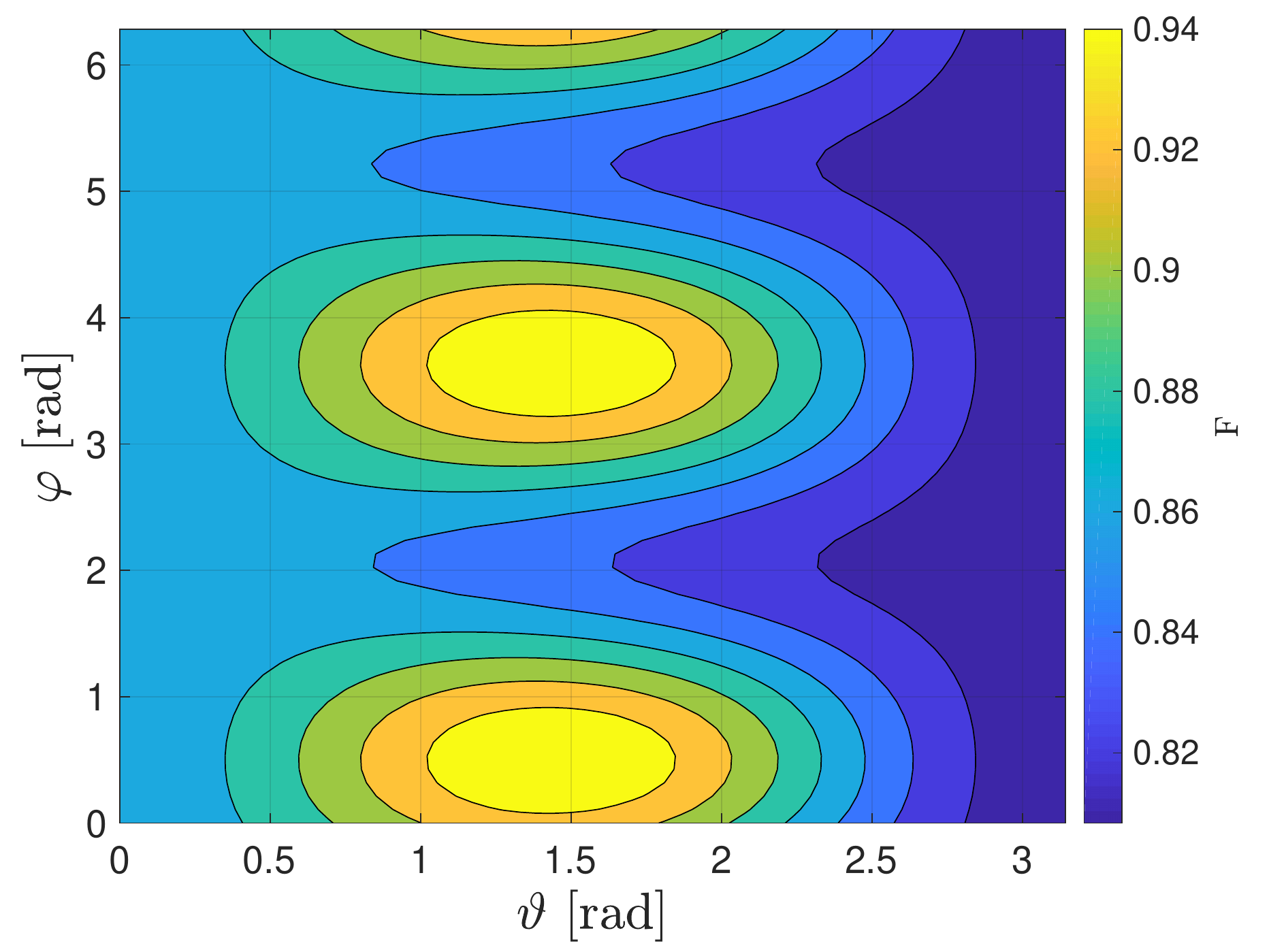}
\caption{Left: for the initial state 
$|\psi\rangle =\sqrt{0.2}|g\rangle+i \sqrt{0.8}|e\rangle$,
fidelity $F$ (full curve, left axis)
and mean photon number $\langle n \rangle$ (right axis, dashed curve) as a 
function of the qubit-cavity coupling strength $g$. 
The mean photon number is also shown for the pure DCE (dotted curve).
Right: Contour plot for fidelity as a function of the Bloch sphere angles
$\theta$ and $\varphi$ for the input state, at $g=0.4$.} 
\label{fig:fidelity}
\end{figure}

The fidelity $F$ as a function of the coupling strength $g$
is shown in Figure~\ref{fig:fidelity} (left plot, full curve), 
for a specific state
$|\psi\rangle$, while the dependence of $F$ on the initial state
can be seen (for a value of $g$ in the USC regime) 
in the right plot of the same figure. 
The state transfer is perfect ($F=1$) in the RWA limit
$g\to 0$. In the ultrastrong coupling regime ($g> 0.1$), $F$ drops
significantly. Moreover, the fidelity is a non-monotonic
function of the coupling strength, 
with maxima at $g^{(M)}_k\approx \omega/(2k+1)$
and minima at $g^{(m)}_k\approx \omega/(2k)$ ($k=1,2,...$; $\omega=1$ in our units).
This regular structure is a
consequence of the terms beyond the RWA in Hamiltonian
(\ref{eq:noREWAquantum}). Indeed, the Bloch vector (of $\textsf{Q}_1$
when $\textsf{Q}_1$ and $\textsf{C}$ are coupled or of $\textsf{Q}_2$
when the interaction is between $\textsf{Q}_2$ and $C$) rotates
with a speed oscillating with frequency $2\omega$ and therefore
also the distance between the exact and the RWA evolution
exhibits oscillations of frequency $2\omega$~\cite{noRWA}. 

The strong connection between the fidelity decay and the DCE is 
witnessed by the fact that the fidelity exhibits anticorrelation of 
peaks and valleys with the mean number $\langle n \rangle$ of photons 
generated in the cavity, both at the end of the protocol 
(dashed curve in the left plot of Figure ~\ref{fig:fidelity}) and for the ``pure'' DCE 
(dot-dashed curve in the same figure). In the latter case, 
qubit $\textsf{Q}_2$ and the cavity $\textsf{C}$ are prepared 
in their ground state and the evolution of 
system $\textsf{Q}_1\textsf{C}$  is followed up to time $\tau$. 
Note that the evolution of a generic input state for $\textsf{Q}_1$
also includes the evolution of the non-interacting ground-state 
$|g\rangle_1|0\rangle$, that is, the pure DCE. 
The photons generated by the pure DCE can lead to further, 
stimulated emission of photons but also to the coherent 
destruction of photons 
(anti-DCE \cite{antiDCE,Motazedifard}).

In the right plot of Figure~\ref{fig:fidelity}, we show the dependence of 
the fidelity for the Rabi transfer protocol on the initial state,
$|\psi\rangle=\cos(\theta/2)|g\rangle+e^{i\varphi}\sin(\theta/2)|e\rangle$,
for $g=0.4$. While in this regime the fidelity is in general 
significantly deteriorated
($F\approx 0.8-0.9$), a non trivial dependence on the Bloch sphere angles 
$\theta$ and $\varphi$ can be seen.

\section{Discussion}

We have shown that the DCE severely limits the performance of 
quantum information protocols in the ultra-strong coupling regime, 
even in the ideal case considered here where dynamics is not affected
by dissipation. 
While the results presented above are for the state transfer protocol,
we have shown in \cite{casimirqip} that more generally the DCE
puts an intrinsic limit to the capability of the
Rabi-based protocols to transmit quantum information.
Novel schemes are required
in order to counteract the DCE.
Preliminary results 
\cite{Catania} with a protocol inspired by stimulated
Raman adiabatic passage (STIRAP)~\cite{STIRAP} 
show enhanced reliability of state transfer up to relatively large
$g\approx 0.2$, also in presence of cavity damping.
Optimal control techniques~\cite{simone,DCEoptimal}
could also be applied to this problem,
with foreseen further improvements for the reliability of quantum 
protocols in the USC regime.

\vspace{6pt} 

\acknowledgments{We acknowledge support by the INFN through the project “QUANTUM”}.

\reftitle{References}



\end{document}